\documentclass[a4paper, 12pt,twoside=false]{scrartcl}
\pdfoutput=1

\usepackage[T1]{fontenc}

\usepackage{dsfont}
\usepackage{savesym}
\usepackage{amsmath, amssymb, amsxtra, bm, scalerel}
\savesymbol{iint}
\usepackage{txfonts}
\restoresymbol{TXF}{iint}

\usepackage[amsmath, amsthm, thmmarks,framed]{ntheorem}
\usepackage[cal=rsfso, calscaled=.96]{mathalfa}
\usepackage{nicefrac}
\usepackage{microtype}
\usepackage{enumitem}
\usepackage{framed}
\usepackage{array}
\usepackage{cite}
\usepackage{todonotes}
\usepackage{mathtools}
\usepackage{wrapfig}
\usepackage[affil-it]{authblk}

\usepackage{hyperref}

\theoremstyle{definition}
\theoremseparator{.}
\newtheorem{defi}{Definition}[section]

\theoremstyle{plain}
\theoremsymbol{}
\newtheorem{lemma}[defi]{Lemma}
\newtheorem{prop}[defi]{Proposition}
\newtheorem{thm}{Theorem}
\newtheorem{coro}[defi]{Corollary}

\theoremstyle{empty}
\theoremheaderfont{\normalfont\upshape\bfseries}
\theorembodyfont{\normalfont\itshape}

\theoremstyle{empty}
\theoremheaderfont{\scshape\bfseries}
\theorembodyfont{\upshape}
\theoremsymbol{\ensuremath{\blacksquare}}
\newtheorem{Proof}{Proof} 			
\def\ben{\begin{equation}}
\def\een{\end{equation}}
\def\non{\nonumber}

\newcommand{\cA}{\mathcal A}

\newcommand{\cD}{\mathcal D}

\newcommand{\cF}{\mathcal F}

\newcommand{\cM}{\mathcal M}
\newcommand{\cN}{\mathcal N}
\newcommand{\cO}{\mathcal O}

\newcommand{\cS}{\mathcal S}

\newcommand{\cU}{\mathcal U}

\newcommand{\cW}{\mathcal W}

\newcommand{\IC}{\mathbb C}
\newcommand{\IM}{\mathbb M}
\newcommand{\IN}{\mathbb N}
\newcommand{\IR}{\mathbb R}

\renewcommand\b{\beta}
        
\renewcommand\d{\delta}        
\newcommand\eps{\varepsilon}

\renewcommand\th{\vartheta}    \newcommand\Th{\Theta}
\newcommand\kap{\kappa}
\renewcommand\l{\lambda}

\newcommand\s{\sigma}

\renewcommand\o{\omega}        \renewcommand\O{\Omega}

		\newcommand{\fw}{\mathfrak{w}}	\newcommand{\sfw}{\mathsf w}

\newcommand{\from}{\colon}

\newcommand{\del}{\partial}

\newcommand{\abs}[1]{{|{#1}|}}			
\newcommand{\eins}{{\mathds{1}}}				
\newcommand{\WF}{{\mathrm{WF}}}				

\newcommand{\vekt}[1]{{\bm{#1}}}

\newcommand{\bb}{{\bm{\b}}}					

\makeatletter
\def\@fnsymbol#1{\ensuremath{\ifcase#1\or  \mathparagraph\or \|\or **\or \dagger\dagger
   \or \ddagger\ddagger \else\@ctrerr\fi}}
\setkomafont{title}{\normalfont\bfseries}
\patchcmd{\@maketitle}{\titlefont\huge}{\titlefont\small}{}{}
\makeatother

\title{\LARGE Spacetime Dependence of Local Temperature in Relativistic Quantum Field Theory}
\author[1,2]{\large Michael Gransee\thanks{Electronic address: gransee [a t] mis. mpg. de}}
\affil[1]{\normalsize MPI f\"ur Mathematik in den Naturwissenschaften, 04103 Leipzig, Germany}
\affil[2]{\normalsize Institut f\"ur Theoretische Physik, Universit\"at Leipzig, 04103 Leipzig, Germany}
\date{\vspace{-5ex}}

\begin{document}
\maketitle

\begin{abstract}\noindent \textbf{Abstract:} The spacetime dependence of the inverse temperature four-vector $\bb$ for certain states of the quantized Klein-Gordon field on (parts of) Minkowski spacetime is discussed. These states fulfill a recently proposed  version of the Kubo-Martin-Schwinger (KMS) boundary value condition, the so-called ``local KMS (LKMS) condition''. It turns out that, depending on the mass parameter $m\geq 0$, any such state can be extended either (i) to a LKMS state on some forward or backward lightcone, with $\bb$ depending linearily on spacetime, or (ii) to a thermal equilibrium (KMS) state on all of Minkowski space with constant $\bb$. This parallels previously known results for local thermal equilibrium (LTE) states of the quantized Klein-Gordon field. Furthermore, in the case of a massless field our results point to a discrepancy with some classic results in general approaches to (non-quantum) relativistic thermodynamics. 
\end{abstract}

\section{Introduction}

It is widely accepted that in the algebraic approach to quantum statistical mechanics global thermal equilibrium states of quantum physical systems can be characterized  by means of the so-called \emph{Kubo-Martin-Schwinger (KMS) condition} \cite{HHW67, BR97, Haa92}. The latter condition is based on  analyticity and periodicity properties of certain correlation functions and the investigation of its consequences led to deep results concerning the expected stability and passivity properties of equilibrium states \cite{HTP77, PW78}. However, the description of non-equilibrium phenomena in quantum systems, and especially in quantum field theory, has not seen such far-reaching progress. This does not come as a surprise, since in view of the wide range of non-equilibrium phenomena the task of describing systems with varying thermal parameters (temperature, pressure, entropy density,...) is extremely complicated even in non-relativistic, non-quantum statistical mechanics, see e.g. \cite{CVJ03}. The subtle point is the question how to define these thermal parameters precisely. Some rigorous results in this direction have been obtained for steady states of non-relativistic quantum systems \cite{Rue00} and in conformal field theory \cite{HL16}, as well as for local thermal equilibrium (LTE) states in non-relativistic quantum systems \cite{Sew13} and LTE states in relativistic quantum field theory on flat and curved spacetimes \cite{BOR02,Buc03,SV08,DHP11,Ver12}.

Another approach towards the description of local equilibrium states in relativistic quantum field theory in the spirit of quantum statistical mechanics and the KMS condition has been recently proposed in \cite{GPV15}. Let us briefly sketch the underlying idea. For simplicitly we consider the quantized Klein-Gordon field on Minkowski spacetime, with field operators formally written as $\phi(x)$. For any (sufficiently regular) state $\o$ of the system one can then define the ``relative-variable correlation function at the spacetime point $q{\ }$'' informally by
\ben
\sfw_{q}(z):=\langle\phi(q-z)\phi(q+z)\rangle_\o, \quad z\in\IR^4,
\een
where $\langle\cdot\rangle_\o$ denotes the expectation value with respect to $\o$. Leaving some technicalities aside (for the precise definition see \cite{GPV15} and Section 3 below) the analyticity properties of the relative-variable correlation function of KMS states give motivation to the following definition:

\textit{A ``sufficiently regular'' state is said to fulfill the \emph{local KMS condition at the spacetime point $q$} if there exists a future-pointing timelike four-vector $\bb$ and a function $F_{q}$ which is holomorphic on the flat tube}
\ben
S_{q}:=\{z+i\s\in\IC^4:\s=\l\bb,0<\l<1\},
\non
\een
\textit{such that $F_{q}$ has boundary values}
\begin{align*}
\lim\limits_{\l\to 0^+} F_{q}(z+i\l\bb)=\sfw_{q}(z) \quad \text{and} \quad \lim\limits_{\l'\to 0^+} F_{q}(z+i(1-\l')\bb)=\sfw_{q}(-z).
\end{align*}
\textit{The latter relations have to be understood in the sense of distributions. Furthermore, $F_{q}$ is assumed to fulfill certain polynomial growth bounds as well as a version of the weak-clustering property \cite{BB96}.} 

As it has been demonstrated in \cite{GPV15}, the LKMS condition (with a proper definition of ``sufficient regularity'') is equivalent to the local thermal equilibrium (LTE) condition of Buchholz, Ojima and Roos \cite{BOR02}. The latter is based on the pointwise comparison of the expectation values $\o(\th(q))$ of certain pointlike localized (thermal) observables $\th(q)$ with their expectation values in (mixtures of) global equilibrium states $\o_{\text{eq}}$. Here, ``sufficiently regular'' means that the states are assumed to fulfill a remnant of the analytic microlocal spectrum condition \cite{SVW02}. Using techniques from microlocal analysis of distributions \cite{Hoe90}, one finds that this assumption imposes stronger analyticity properties on the correlation functions than those imposed by the LKMS condition alone.  In this respect, the LKMS condition above may be viewed as a (covariant) remnant of the \emph{relativistic KMS condition}  \cite{BB94} for certain non-equilibrium states of the quantized Klein-Gordon field on patches of Minkowski spacetime. For a slightly less technical review of the LTE and LKMS condition and their interrelation, we also refer to \cite{Gra16}.

Having established such a relation a natural question arises: If a state fulfills the LKMS condition at all points of a region $\cO$, what can we say about the possible spacetime dependence of the inverse-temperature four-vector field $\bb\from\cO\ni q\mapsto \bb(q)\in V_+$? Skipping the technical details, which will be presented below, the answer to this question is the following:

\textit{For the massless quantized Klein-Gordon field either (1) $\bb=\text{const.}$,  and the state can be extended to a KMS state on all of Minkowski spacetime or (2) $\bb(q)=cq+\tilde{\bb}$ with constants $c\in\IR$ and $\tilde{\bb}\in\IR^4$ , and the state can be extended to an LKMS state defined in some open (forward or backward) lightcone.}

\textit{For the massive quantized Klein-Gordon field $\bb$ is constant and the state can be extended to a KMS state on all of Minkowski spacetime.}

We first note that these results are in accordance with similar results obtained by implementing the LTE condition for the quantized Klein-Gordon field \cite{Buc03, Hue05}. In particular, Buchholz has shown that for LTE states with varying thermal parameters the maximal region of thermaliticity is always contained in a timelike simplicial cone, i.e. an intersection of characteristic half-spaces, cf. \cite[Prop. 5.2]{Buc03}.  The above result demonstrates that, as a consequence of the analytic Hadamard condition, the LKMS condition is more restrictive than the LTE condition. Namely, for LKMS states with varying $\bb$ it leads to a fixed maximal LKMS region, given by some open (forward or backward)  lightcone. On the one hand, this is not surprising in view of our additional analyticity assumption on the correlation functions. On the other hand, we do not have to make additional assumptions on the form of the initial LKMS region $\cO$, in contrast to \cite[Prop. 5.2]{Buc03} where the initial region of thermaliticity is already assumed to include some lightcone.

Let us also compare the above result with classic results on the spacetime dependence of the (inverse) temperature in non-quantum relativistic thermodynamics, cf. \cite{Dix78}. We denote the \emph{entropy current density} by $s^\mu\in V_+$. The \emph{entropy source strength} $\sigma$ is defined as the four-divergence of $s^\mu$, i.e. $\s:=\del_\mu s^\mu$. The second law of thermodynamics is then expressed by\footnote{Usually, the second law of thermodynamics is assumed to hold only for isolated systems. Dixon \cite[\S 2 in Ch. 4]{Dix78} gives an argument that this requirement is not relevant for practical purposes and that eq.\ \eqref{eq:secondlaw} should therefore be valid for essentially all physical systems.}
\ben
\sigma\geq0.
\label{eq:secondlaw}
\een
Making some physically motivated assumptions on the functional dependence of the entropy current density $s^\mu$ on the stress-energy and matter distribution of the system one finds that the second law of thermodynamics ($\s=0$) implies that for the global equilibrium states of a relativistic simple fluid it holds
\ben
\bb_\nu(q)=C_{\mu\nu}q^\mu+\tilde{\bb}_\nu,
\label{eq:Dixonresult}
\een
where $C_{\mu\nu}$ is a constant $\binom{0}{2}$-Lorentz tensor with $C_{\nu\mu}=-C_{\mu\nu}$, and $\tilde{\bb}\in\IR^4$ is a constant four-vector. It is implicitly understood that eq.\ \eqref{eq:Dixonresult} is physically meaningful only for those $q\in\IM$ for which $\bb(q)$ is future-pointing timelike. It is immediately clear that our results for the massless Klein-Gordon field, expressed above (and more precisely in Theorem \ref{thm:mainresult} in Section 3 below), are in conflict with eq.\ \eqref{eq:Dixonresult}. It seems to be likely that the inconsistency of eq.\ \eqref{eq:Dixonresult} and our results for the \emph{massless} Klein-Gordon field originate from the fact that the situation considered in \cite{Dix78} refers to a classical ideal fluid which is assumed to be \emph{massive} (i.e. it has positive mass-energy density everywhere). This point of view is supported by the observation that our results for the massive Klein-Gordon field are not in conflict with eq.\ \eqref{eq:Dixonresult}. In fact, the statement of Proposition \ref{prop:betaofq} for $m>0$ is just a special case of the latter, namely $C_{\mu\nu}\equiv 0$. 
It seems to be worthwhile to investigate this seeming discrepancy between the macroscopic and microscopic theory in more detail in order to gain more insight into the relativistic non-equilibrium thermodynamics of massless quantum fields.

The article is organized as follows: In Section 2 we fix our framework by collecting the necessary notation and definitions. In Section 3 we give a precise definition of the LKMS condition and state our main result. Section 4 is devoted to the proof of the main result, while Section 5 contains some concluding remarks. The Appendix contains the proofs of two technical lemmas needed in the proof of the main result, Theorem \ref{thm:mainresult}.

\section{General framework}

As usual, the space of Schwartz functions on $\IR^n$ is denoted by $\cS(\IR^n)$, and the space of smooth functions with compact support contained in some open subset $X\subseteq \IR^n$ is denoted by $\cD(X)$. We write $\cD'(X)$ for the space of distributions on $X$, and $\cS'(\IR^n)$ for the space of tempered distributions on $\IR^n$.  

We denote Minkowski spacetime by $\IM=(\IR^4,\eta)$ with the usual Lorentzian pseudo-metric $\eta_{\mu\nu}=\text{diag}(1,-1,-1,-1)$. We implicitly assume summation over repeated indices (Einstein summation convention). For the remainder of this work $\cM\subseteq\IM$ denotes a globally hyperbolic, open and convex subregion of Minkowski spacetime. The open forward lightcone $V_+$ is defined by
\ben
V_+:=\{e\in T\IM:e^\mu e_\mu>0,e_0>0\}.
\non
\een
The boundary of $V_+$ is denoted by $\del V_+$. Th open backward lightcone is defined as $V_ -:=-V_+$ and its boundary is denoted by $\del V_ -$. The open lightcones emanating from a point $q\in\IM$ are defined by 
\ben
V_\pm(q)=V_\pm+q:=\{v\in\IR^4:v-q\in V_\pm\}.
\non
\een
The partial derivatives with respect to the coordinate $x^\mu$ are denoted by $\del_\mu:=\del/\del x^\mu$. The D'Alembert operator is defined by $\Box:=\del_\mu\del^\mu$.

The Fourier transform of a function $f$ is symbolically written as $\cF\left[f(x)\right](p)$ or $\hat{f}(p)$ in 4 dimensions and $\cF\left[f(\vec{x})\right](\vec{p})$ or $\hat{f}(\vec{p})$ in 3 dimensions, respectively. Our conventions are
\begin{align*}
\cF\left[f(x)\right](p)&=\frac{1}{4\pi^2}\int\limits_{\IR^4}d^4x\ f(x)e^{ip^\mu x_\mu}\quad \text{and} \\[1ex]
\cF\left[f(\vec{x})\right](\vec{p})&=\frac{1}{\sqrt{2\pi}^{3/2}}\int\limits_{\IR^3}d^3\vec{x}\ f(\vec{x})e^{-i\vec{x}\cdot\vec{p}},
\end{align*} 
where $\vec{x}\cdot\vec{p}$ denotes the Euclidean inner product of $\vec{x},\vec{p}\in\IR^3$. The same symbolic notation applies for tempered distributions.

The field $\phi(x)$ is thought of as an operator valued distribution and the field operators $\phi(f)$, formally given by $\phi(f)=\int\phi(x)f(x)d^4 x$ for $f\in\cD(\cM)$, fulfill the Klein-Gordon equation in the sense of distributions,
\ben
\phi\left((\Box+m^2)f \right)=0\quad \forall f\in\cD(\cM).
\non \een
Formally, the latter can be written as
\ben
(\Box+m^2)\phi(x)=0\quad x\in\cM.
\label{eq:KGformal}
\een
The field is regarded to be Hermitian, i.e. $\phi(f)^*=\phi(\bar{f})$ for all $f\in\cD(\cM)$ and the field operators are supposed to fulfills the canonical commutation relations (CCR):
\ben
[\phi(f),\phi(g)]=iC(f,g)\eins\quad \forall f,g\in\cD(\cM),
\non \een
where $[\phi(f),\phi(g)]:=\phi(f)\phi(g)-\phi(g)\phi(f)$ is the commutator of $\phi(f)$ and $\phi(g)$. The (globally defined) distribution $C\in\cS'(\IR^4\times\IR^4)$ is thus called the \emph{commutator function}\footnote{Sometimes $C$ is called the \emph{causal propagator} in view of the fact that it can be defined as the advanced minus retarded fundamental solution of the Klein-Gordon equation.}. The commutator function is translation invariant, $C(x,y)\equiv C(x-y)$, and is given via its Fourier transform
\ben
\widehat{C}(p)=-\frac{i}{2\pi}\eps(p_0)\d(p^2-m^2),
\label{eq:spectrfunc}
\een 
where $\d$ denotes the usual Dirac delta distribution and $\eps$ is the (generalized) sign function, i.e. $\eps(p_0)=p_0/\abs{p_0}$. The field algebra of the Klein-Gordon field with mass parameter $m\geq 0$ on $\cM$ is then is (abstract) unital ${}^*$-algebra $\cA_m(\cM)$, generated by the unit $\eins$ and polynomials in the field operators $\phi(f)$. 
In this way we obtain an (abstract) quantization of the Klein-Gordon field in terms of (abstract) operators $\phi(f)$.  A concrete realization of $\cA_m(\cM)$ can be given in terms of the Borchers-Uhlmann algebra over tensor products of test functions, cf. \cite{Haa92}.

A \emph{state on} $\cA_m(\cM)$ is a positive linear functional $\o\from\cA_m(\cM)\to\IC$. Its $2$-\emph{point function} is the distribution $\o_2\in\cD'(\cM\times\cM)$ which is given by
\ben
\o_2(f,g)=\o(\phi(f)\phi(g))\quad \forall f,g\in\cD(\cM).
\non \een
or in a more formal notation,
\ben
\o_2(x,y)=\o(\phi(x)\phi(y)),\quad x,y\in\cM.
\non \een 
Symbolically, it follows from eq.\ \eqref{eq:KGformal} that the two-point function $\o_2$ fulfills the Klein Gordon equation in both its entries: 
\ben
(\Box_x+m^2) \o_2(x,y)=(\Box_y+m^2)\o_2(x,y)=0,
\label{eq:twoKG}
\een
which has to be interpreted in the sense of distributions, i.e.
\ben
\o_2(\Box f, g)=\o_2(f,\Box g)=-m^2\o_2(f,g)\quad \forall f,g\in\cD(\cM).
\label{eq:KGdistr}
\een

For notational simplicity in the upcoming discussion we will restrict ourselves to the class of \emph{quasifree} (or \emph{Gaussian}) \emph{states}: A quasifree state $\o$  on $\cA_m(\cM)$ is completely determined by its two-point function $\o_2$ via the relation
\ben
\o\left(e^{it\phi(f)}\right)=\exp\left(-\frac{1}{2}\o_2(f,f)\cdot t^2\right),
\non \een
to be understood in the sense of formal power series in $t$.

We will also restrict ourselves to the class of \emph{analytic Hadamard states} WHICH are characterized by certain regularity properties of their two-point function $\o_2$. Following \cite{SVW02} we will encode these in terms of the analytic wave front set \cite{Hoe90} of $\o_2$:
\begin{defi}
\label{defi:Hadamard}
A quasifree state $\o$ on $\cA_m(\cM)$ is called an \emph{analytic Hadamard state} iff
\ben
 \WF_A(\o_2)=\{(x,x^\prime;k,-k)\in T^\ast\cM^2\backslash\{0\}:x\sim_k x^\prime,k_0>0\}.
\label{eq:WFAmink}
\een
Here $x\sim_k x^\prime$ means that $x$ and $x'$ can be connected by the uniquely defined lightlike geodesic (a straight line in $\cM$) with cotangent vector $k$, while the condition $k_0>0$ requires $k$ to be future directed. 
\end{defi}
The most important consequence of Definition \ref{defi:Hadamard} in the present context is that the two-point function $\o_2(\cM)$ of any analytic Hadamard state $\o$ on $\cA_m(\cM)$ has a singularity behaviour which mimicks that of $\o_2^{\text{vac}}$, the two-point function of the unique vacuum state $\o_{\text{vac}}$ on $\cA_m(\IM)$. In particular, $\o_{\text{vac}}$ itself fulfills the analytic Hadamard condition. Moreover, let us define $\cW$, the \emph{regular part} of the two-point function $\o_2\in\cD'(\cM\times\cM)$, as the difference 
\ben
\cW:=\o_2-\left.\o_2^\text{vac}\right|_{\cM\times\cM},
\non \een
where $\left.\o_2^\text{vac}\right|_{\cM\times\cM}\in\cD'(\cM\times\cM)$ is the restriction of $\o_2^\text{vac}$ to $\cD(\cM\times\cM)$. Then the analytic Hadamard condition, eq.\ \eqref{eq:WFAmink}, implies that $\cW\in C^A(\cM\times\cM)$, where $C^A(X)$ denotes the class of real-analytic functions on an open subset $X\subseteq\IR^4$. As discussed in \cite{GPV15}, this analyticity property of $\cW$ plays a key role in establishing the  equivalence of the LKMS condition and the LTE condition of Buchholz, Ojima and Roos, mentioned in the introduction. For a detailled account of the (analytic) Hadamard condition in the context of quantum field theory in curved spacetimes as well as its physical motivation, the reader is referred to \cite{Wal94,Rad96,SVW02,FV13,Hac15} and references therein.

Next, we would like to introduce a certain set of coordinates on $\cM\times\cM$ which is adapted to our problem.  For this we first consider the following subset of $\cM\times\IR^4$:
\begin{align*}
\cN&:=\{(q,z)\in\cM \times\IR^4:q\pm\frac{z}{2}\in\cM\}.
\end{align*}
and define the coordinate transformation (diffeomorphism) $\kap\from\cN\to\cM\times\cM$ via
\ben
\kap(q,z)\equiv\left(x(q,z),y(q,z)\right):=\left(q-\frac{z}{2},q+\frac{z}{2}\right)\quad \forall (q,z)\in\cN.
\label{eq:kappa}
\een
The inverse transformation $\kap^{-1}\from\cM\times\cM\to\cN$ is given by
\ben
\kap^{-1}(x,y)\equiv(q(x,y),z(x,y))=\left(\frac{x+y}{2},y-x\right)\quad \forall x,y\in\cM.
\non \een
We call $(q,z)$ the \emph{point-split coordinates}. Furthermore, we call $q$ the \emph{position variable} and $z$ is called the \emph{relative variable}. 

As shown in \cite[Lemma 3.1]{GPV15} the formal expression
\ben
{\mathsf w}_q(z):=\o_2\left(q-\frac{z}{2},q+\frac{z}{2}\right)
\non \een
can be meaningfully defined as a distribution in $\cD'(\cU_{q})$ for any \emph{fixed} $q\in\cM$, where $\cU_{q}\subseteq\IR^4$ is the open neighbourhood of $0\in\IR^4$ defined by
\ben
\cU_{q}:=\{z\in\IR^4:(q,z)\in\cN\}.
\label{eq:Uq}
\een
The proof of this is based on H\"ormander's criterion (cf. \cite[Thm. 8.2.4 and Thm. 8.5.1]{Hoe90}. By applying analogous arguments we obtain a similar result for the case of a varying position variable $q$:

\begin{lemma}
\label{lemma:pullback}
Let $\o$ be an analytic Hadamard state on $\cA_m(\cM)$ and let $\kap \from\cN \to\cM\times\cM$ be defined by eq \eqref{eq:kappa}. Then ${\mathsf w}_2:=\kap^*\o_2$, the pullback of $\o_2$ with respect to $\kap$, can be defined as a distribution in $\cD'(\cN)$, such that
\ben
\WF_A({\mathsf w}_2)\subset \{(q,z;0,k)\in T^*\cN\backslash\{0\}:k_\mu=\mathrm{sign}(z_0)z_\mu, z\in\del V_+(q)\cup\del V_-(q)\}.
\label{eq:WFAw}
\een
\end{lemma}

\begin{Proof}[Proof.]
The set of normals of the map $\kap$ is defined by
\ben
N_\kap=\{(\kap(q,z),\eta)\in (\cM\times\cM)\times(\IR^4\times\IR^4):(\kap')^T(q,z)[\eta]=0\}.
\non \een
Since $\kap$ is linear we have $\kap'(q,z)=(J\kap)(q,z)$, with a constant Jacobi matrix $J\kap$. Writing $\eta=\begin{pmatrix}
\eta_1\\
\eta_2
\end{pmatrix}$, we find that \ben
(J\kap)^T(q,z)[\eta]=\begin{pmatrix}
					\eins_4 & \eins_4\\
					-\frac{1}{2}\eins_4& \frac{1}{2}\eins_4
					\end{pmatrix}[\eta]
					=\begin{pmatrix}
					\eta_1+\eta_2\\
					\frac{1}{2}(\eta_2-\eta_1)
					\end{pmatrix}
\non \een
which vanishes if and only if $\eta_1=\eta_2=0$. Thus, we have
\ben
N_\kap=\{(q-z/2,q+z/2,0,0):(q,z)\in\cN\}.
\non \een
Comparing this with the analytic wave front set $\WF_A(\o_2)$, given by eq.\ \eqref{eq:WFAmink}, we see that $N_\kap\cap\WF_A(\o_2)=\emptyset$. Since $\kap$ is a real-analytic map we can apply Theorem 8.5.1 of \cite{Hoe90} to find that $\kap^*\o_2\in\cD'(\cN)$ is uniquely defined and that eq.\ \eqref{eq:WFAw} holds. 
\end{Proof}
In the following, for any Hadamard state $\o$ on $\cA_m(\cM)$ we will call ${\mathsf w}_2=\kap^* \o_2$ the \emph{point-split variable two-point function of} $\o$. We note that eq.\ \eqref{eq:WFAw} reveals that the singularity behaviour of $\sfw_2$ is basically encoded in the relative variable $z$. 

It is clear that the point-split variable two-point function ${\mathsf w}_2\in\cD'(\cN)$ contains the same information on the state $\o$ as is contained in the two-point function $\o_2\in\cD'(\cM\times\cM)$. In particular, any quasifree state is completely determined by ${\mathsf w}_2$. However, the choice of the point-split coordinates seems to be more convenient when it comes to the desription of local properties of the state $\o$ in some open neighbourhood $\cO\subset \cM$. To proceed,  for any open convex neighbourhood $\cO\subseteq\cM$ we define the open subsets $\cN_\cO\subseteq\cO\times\IR^4$ via
\ben
\cN_\cO:=\{(q,z)\in\cO\times\IR^4:q\pm\frac{z}{2}\in\cO\},
\non \een
The local properties (e.g. of thermal nature) will be encoded by giving respective conditions only for the restriction of ${\mathsf w}_2$ to $\cD(\cN_\cO)$ and we will denote the latter by
\ben
{\mathsf w}_\cO:=\left.{\mathsf w}_2\right|_{\cN_0}\in\cD'(\cN_\cO).
\non \een
We will occasionally call ${\mathsf w}_\cO$ the \emph{point-split two-point function of $\o$ in $\cO$.} 

\section{Main result}

As discussed in the introduction, the local KMS (LKMS) condition at a spacetime point $q\in\cM$, given in precise terms in \cite{GPV15}, is to be seen as a remnant of the relativistic KMS (rKMS) condition of \cite{BB94}. The latter is based on analyticity and periodicity properties of position space correlation functions.  In \cite{BB96} an equivalent formulation of the rKMS condition in Fourier (momentum) space in terms of spectral properties of these states. For our purposes it turns out to be most convenient to rely to a similar spectral formulation of the (pointwise) LKMS condition which has been shown to be equivalent to its position space counterpart in \cite[Thm. 3.8]{GPV15}. We extend this Fourier space formulation to open spacetime regions $\cO$ as follows:

\begin{defi}[${[\bb,\cO]}$-LKMS condition]
\label{defi:LKMS}
Let $\o$ be an analytic Hadamard state on $\cA_m(\cM)$ and let $\cO\subseteq\cM$ be open and convex.
  We say that $\o$ fulfills the \emph{local KMS condition in $\cO$ with respect to the inverse-temperature four-vector field $\bb$}, or $[\bb,\cO]$-\emph{LKMS condition} for short, iff there exists a map $\bb\in C^2(\cO,V_+)$ such that ${\mathsf w}_\cO\in\cD'(\cN_\cO)$, the point-split two-point function of $\o$ in $\cO$, has an extension $\fw_\cO\in\cD'(\cO\times\IR^4)$, with
\begin{itemize}
\item[(i)] $\fw_\cO(q,\cdot)\in\cS'(\IR^4)$ for all $q\in\cO$.
\item[(ii)] $\fw_\cO(q,t\bb(q))\xrightarrow[\abs{t}\rightarrow \infty]{} 0$ for all $q\in\cO$.
\item[(iii)] For any $q\in\cO$ the following holds in the sense of distributions (w.r.t. the variable $p$):
\ben
e^{\bb(q)p}\widehat{\fw}_\cO(q;-p)=\widehat{\fw}_\cO(q;p),
\non \een
where $\widehat{\fw}_\cO\in\cD'(\cO\times\IR^4)$ denotes the partial Fourier transform of $\fw_\cO$ with respect to the relative variable $z$. 
\end{itemize}
\end{defi}
For a justification of the assumption that $\sfw_\cO$ can be extended to a (tempered) distribution in the relative variable $z$ and the connection to the analytic Hadamard property, see \cite{GPV15}. Furthermore, the assumption (ii) in Def. \ref{defi:LKMS} is to be viewed as a remnant of the assumed time-clustering property of KMS states, cf. \cite{BB96,GPV15}. The role of these cluster-properties is to rule out phase transitions. For global equilibrium states (KMS) this means that for any inverse-temperature-four vector $\bb\in V_+$ there exists a unique state $\o_\bb$ which fulfills the (relativistic) KMS condition with respect to $\bb$. 

The $[\bb,\cO]$-LKMS condition implies by standard arguments \cite{BB96,GPV15}, exploiting the time-clustering property $(ii)$ in Def. 3.1, that the partial Fourier transform of $\fw_\cO$ with respect to the relative variable $z$ is given by
\ben
\widehat{\fw}_\cO(q;p)=\frac{i\widehat{C}(p)}{1-e^{\bb(q)p}},
\non \een
in the sense of distributions on $\cO\times\IR^4$. Here, $$\widehat{C}(p)=-i\cdot\left(\widehat{\fw}_\cO(q;p)-\widehat{\fw}_\cO(q;-p)\right)$$ is the spectral function defined in eq.\ \eqref{eq:spectrfunc}\footnote{Since the commutator function $C$ is translation-invariant, the spectral function $\widehat{C}$ is independent of $q$ and thus independent of $\cO$.}. Thus, for any $[\bb,\cO]$-LKMS state $\o$, the extended point-split variable two-point function $\fw_\cO\in\cD'(\cO\times\IR^4)$ is formally given by the inverse (partial) Fourier transform 
\ben
\fw_\cO(q,z)=\frac{1}{(2\pi)^3}\int d^4 p \frac{\eps(p_0)\d(p^2-m^2)}{1-e^{-\bb(q)p}}e^{-ipz}.
\label{eq:wexpl}
\een
 
The main result of this article concerns the possible spacetime dependence of the inverse-temperature four-vector $\bb$ as well as the maximal regions in which $[\bb,\cO]$-LKMS states with non-constant $\bb$ can exist. It is stated in precise terms as follows:

\begin{thm}
\label{thm:mainresult}
Let $\cO\subseteq\cM$ be open and convex and let $\o$ be an analytic Hadamard state on $\cA_m(\cM)$ which fulfills the $[\bb,\cO]$-LKMS condition. Then the following holds:

\vspace{12pt}
\noindent If $m=0$, there exists a $\tilde{\bb}\in\IR^4$ such that either:
\begin{itemize}
\item[(i)] $\o$ can be extended to a $[\bb,V_+-\tilde{\bb}]$-LKMS state $\widetilde{\o}$ on $\cA_0(V_+-\tilde{\bb})$, with
\ben
\bb(q)=c_\o (q+\tilde{\bb})\quad\forall q\in V_+-\tilde{\bb},
\non \een
where $c_{\o}>0$ is a state dependent constant, or
\item[(ii)] $\o$ can be extended to a $[\bb,V_--\tilde{\bb}]$-LKMS state $\widetilde{\o}$ on $\cA_0(V_--\tilde{\bb})$, with
\ben
\bb(q)=c_\o (q+\tilde{\bb})\quad\forall q\in V_--\tilde{\bb}\quad \forall q\in V_--\tilde{\bb},
\non \een
where $c_{\o}<0$ is a state dependent constant, or
\item[(iii)] $\o$ can be extended to a $\tilde{\bb}$-KMS state $\o_{\tilde{\bb}}$ on $\cA_0(\IM)$.  
\end{itemize}

\noindent If $m>0$, the following are equivalent:
 \begin{itemize}
  \item[(i)] $\o$ is a $[\bb,\cO]$-LKMS state.
 \item[(ii)] There exists a $\tilde{\bb}\in V_+$ such that $\o$ can be extended to a $\tilde{\bb}$-KMS state $\o_{\tilde{\bb}}$ on $\cA_{m}(\IM)$.
 \end{itemize}
\end{thm}

As an immediate corollary we obtain the equivalence of LKMS condition and the KMS condition for globally defined states of the free scalar field on Minkowski spacetime:

\begin{coro}
Let $\o$ be an analytic Hadamard state on $\cA_m(\IM)$, with $m\geq 0$. Then the following are equivalent:
\begin{itemize}
\item[(i)] $\o$ fulfills the $[\bb,\IM]$-LKMS condition.
\item[(ii)] $\o$ is a $\tilde{\bb}$-KMS state for some $\tilde{\bb}\in V_+$.
\end{itemize}
\end{coro}

\section{Proof of the main result}

We divide the proof of Theorem \ref{thm:mainresult} into three steps. First, we derive certain dynamical constraints on the point-split two-point function ${\mathsf w}_2$. Those arise from the equations of motion, i.e. from the Klein-Gordon equation \eqref{eq:twoKG} fulfilled by the two-point function $\o_2$. After these preparations we derive the spacetime dependence of the inverse-temperature four-vector field $\bb$ in the ``LKMS region'' $\cO\subseteq \cM$. The analyticity properties of $\sfw_2$ imposed by the analytic Hadamard condition finally yield the desired extensions of the state $\o$ either to the lightcone regions ocurring in Thm. \ref{thm:mainresult} in the case of variable $\bb$ or to all of Minkowski spacetime in the case of constant $\bb$.  

\subsection{Constraints from the equation of motion}

We first derive certain dynamical constraints on the point-split two-point function ${\mathsf w}_2\in\cD'(\cN)$ which arise from the Klein-Gordon equation, eq.\ \eqref{eq:KGformal}. These are given by the following lemma.

\begin{lemma}
Let $\o$ be an analytic Hadamard state on $\cA_m(\cM)$. Then the following hold in the sense of distributions:
\begin{gather}
 \del_{[q]}^\mu \del_\mu^{[z]}{\mathsf w}_2(q,z)=0,
 \label{eq:relconstr1}\\
\Box_q {\mathsf w}_2(q,z)=-4\left(\Box_z +m^2\right){\mathsf w}_2(q,z).
\label{eq:relconstr2}
\end{gather}
\end{lemma}

\begin{Proof}[Proof.]
We will make use of the well-known fact that the tensor product space $\cD(\cM)\otimes_\text{alg}\cD(\cM)$, where $\otimes_\text{alg}$ denotes the algebraic tensor product, is sequentially dense in $\cD(\cM\times\cM)$ with respect to the usual topology which makes $\cD(\cM\times\cM)$ a locally convex space (see e.g. \cite{RS80}). Thus, it is sufficient to restrict ourselves to linear combinations $h$ of finitely many elements $h_j\in\cD(\cM\times\cM)$ which are of the form $h_j=f_j\otimes g_j$ for some $f_j,g_j\in\cD(\cM)$, i.e.
\ben
h(x,y)=\sum\limits_{j=1}^n(f_j\otimes g_j)(x,y)=\sum\limits_{j=1}^n f_j(x)g_j(y)\quad\forall x,y\in\cM,
\non \een
where $n\in\IN$. If $\kap$ denotes the point-splitting map defined by eq.\ \eqref{eq:kappa}, it follows that the set
\ben
\{h\in\cD(\cN):h(q,z)=\sum\limits_j(f_j\otimes g_j)\left(\kap(q,z)\right) \text{for some } f_j,g_j\in\cD(\cM)\}
\non \een
is dense in $\cD(\cN)$.

By elementary calculations we obtain
\begin{align}
\del_\mu^{[q]}\del^\mu_{[z]}\left[(f\otimes g)\left(\kap(q,z)\right)\right]&=-\frac{1}{2}\Big(\Box f\otimes g-f\otimes \Box g\Big)\left(\kap(q,z)\right),
\label{eq:testconstr1}\\[1ex]
\Box_q\left[(f\otimes g)\left(\kap(q,z)\right)\right]&=\left(\Box f\otimes g +2\del_\mu f\otimes \del^\mu g+f\otimes\Box g\right)\left(\kap(q,z)\right),\quad \mathrm{and}
\label{eq:testconstr2}\\[1ex]
 \Box_z\left[(f\otimes g)\left(\kap(q,z)\right)\right]&=\frac{1}{4}\left(\Box f\otimes g -2\del_\mu f\otimes \del^\mu g+f\otimes\Box g\right)\left(\kap(q,z)\right)\quad \forall (q,z)\in\cN.
\label{eq:testconstr3}
\end{align}
where $\del_{[q]}^\mu$ (resp. $\del_\mu^{[z]}$) denotes differentiation with respect to the position variable $q$ (resp. the relative variable $z$) and $\Box_q, \Box_z$ are the respective D'Alembert operators.
For $h\in\cD(\cM\times\cM)$ the pullback $\kap^*h\in\cD(\cN)$ is given by ordinary composition of maps:
\ben
(\kap^*h)(q,z):=h\left(\kap(q,z)\right)=h\left(q-\frac{z}{2},q+\frac{z}{2}\right)\quad \forall (q,z)\in\cN.
\non \een
Using the linearity of $\o_2$ andthe identity ${\mathsf w}_2\left(\kap^*(f\otimes g)\right)=\o_2(f,g)$ for $f,g\in\cD(\cM)$, we find that the Klein-Gordon equation \eqref{eq:KGdistr} implies the relation
\begin{align}
{\mathsf w}_2\left(\del_\mu^{[q]}\del^\mu_{[z]}\kap^*(f\otimes g)\right)&\stackrel{\eqref{eq:testconstr1}}{=}\frac{1}{2}\left(\o_2\left( f,\Box g\right)-\o_2\left(\Box f,  g\right)\right)\non\\
&=\frac{1}{2}m^2\left(\o_2(f,g)-\o_2(f,g)\right)\non\\
&=0\qquad\forall f,g\in\cD(\cM).
\label{eq:tensconstr}
\end{align}
Since the set of finite linear combinations of functions of the form $f\otimes g$ is dense in $\cD(\cN)$, continuous extension of eq.\ \eqref{eq:tensconstr} yields the first dynamical constraint in Lemma \ref{lemma:constraints}, eq.\ \eqref{eq:relconstr1}. 

Similarily, the Klein-Gordon equation implies the relation
\begin{align}
{\mathsf w}_2\left(\Box_z\kap^*(f\otimes g)\right)&\stackrel{\eqref{eq:testconstr3}}{=}\frac{1}{4}\o_2\left(\Box f,g\right)-\frac{1}{2}\o_2\left(\del_\mu f,\del^\mu g\right)+\frac{1}{4}\o_2\left(f,\Box g\right)\non\\
&=-\frac{1}{2}m^2\o_2(f,g)-\frac{1}{2}\o_2\left(\del_\mu f,\del^\mu g\right),
\label{eq:boxztest}
\end{align}
and, consequently, we have
\begin{align*}
{\mathsf w}_2\left(\Box_q\kap^*(f\otimes g)\right)&\stackrel{\eqref{eq:testconstr2}}{=}\o_2\left(\Box f,g\right)+2\o_2\left(\del_\mu f,\del^\mu g\right)+\o_2\left(f,\Box g\right)\non\\
&=-2m^2\o_2(f,g)+2\o_2\left(\del_\mu f,\del^\mu g\right)\non\\
&\stackrel{\eqref{eq:boxztest}}{=}-4\left({\mathsf w}_2 \left(\Box_z\kap^*(f\otimes g)\right)+m^2{\mathsf w}_2\left(\kap^*(f\otimes g)\right)\right)\quad \forall f,g\in\cD(\cM).
\end{align*}
By continuity this yields the second dynamical constraint in Lemma \ref{lemma:constraints}, eq.\ \eqref{eq:relconstr2}. 
\end{Proof}

In particular, the dynamical constraints \eqref{eq:relconstr1} and \eqref{eq:relconstr2} hold for the unique vacuum state $\o_\text{vac}$ on $\cA_m(\IM)$. We denote by  ${\mathsf w}_2^\text{vac}\in\cD'(\cN)$ the point-split two-point function of the restricted state $\left.\o_\text{vac}\right|_\cM$. Then for any analytic Hadamard state $\o$ on $\cA_m(\cM)$ the regular part $\mathcal{W}_2\in C^A(\cN)$ of ${\mathsf w}_2\in\cD'(\cN)$, defined by $\mathcal{W}_2(q,z):=({\mathsf w}_2-{\mathsf w}_2^\text{vac})(q,z)$ is subject to the following constraints:
\begin{gather}
  \del_{[q]}^\mu \del_\mu^{[z]}\mathcal{W}_2(q,z)= 0,\ \textrm{and}
  \label{eq:Wconstr1} \\[1ex]
  \Box_q \mathcal{W}_2(q,z)=-4\left(\Box_z+m^2\right)\mathcal{W}_2(q,z),\quad \forall (q,z)\in\cN.
\label{eq:Wconstr2}
\end{gather}

\subsection{Spacetime dependence of inverse temperature}

We next show how the dynamical constraints given by eq. \eqref{eq:Wconstr1} and eq.\  \eqref{eq:Wconstr2} lead to constraints on the inverse temperature four-vector field $\bb\in C^2(\cO,V_+)$ of a $[\bb,\cO]$-LKMS state $\o$ on $\cA_m(\cM)$. 
According to the discussion following Definition \ref{defi:Hadamard}, the singular part of the distribution $\fw_\cO$ is given by $\fw_\cO^\text{vac}\in\cD'(\cO\times\IR^4)$, the point-split two-point function  of the unique vacuum state $\o_\text{vac}$ in $\cO$, which is $q$-independent and is formally given by
\ben
\fw_\cO^\text{vac}(q,z)=\frac{1}{(2\pi)^3}\int d^4 p\ \Th(p_0)\d(p^2-m^2)e^{-ipz}.
\non \een
Using eq.\ \eqref{eq:wexpl} we thus find that the regular part $\mathcal{W}_\cO\in C^A(\cO\times\IR^4)$ of $\fw_\cO$ is given by
\begin{align*}
\mathcal{W}_\cO(q,z)&:=(\fw_\cO-\fw_\cO^\text{vac})(q,z)\non\\[1ex]
&=\frac{1}{(2\pi)^3}\int d^4 p \left(\frac{\eps(p_0)\d(p^2-m^2)}{1-e^{-\bb(q)p}}-\Th(p_0)\d(p^2-m^2)\right)e^{-ipz}\quad \forall q\in\cO, z\in\IR^4.
\end{align*}
Integrating over the $p_0$-component yields
\begin{align}
\mathcal{W}_\cO(q,z)&= \frac{1}{(2\pi)^3}\int_{\IR^3} \frac{d^3\vec{p}}{2\o_{\vec{p}}}\left(\frac{e^{-i\o_{\vec{p}}t}e^{i\vec{p}\vec{z}}}{1-e^{-\b_q\o_{\vec{p}}}e^{\vec{\b}(q)\vec{p}}}-\frac{e^{i\o_{\vec{p}}t}e^{i\vec{p}\vec{z}}}{1-e^{\b_q\o_{\vec{p}}}e^{\vec{\b}(q)\vec{p}}}-e^{-i\o_{\vec{p}}t}e^{i\vec{p}\vec{z}}\right)\non\\[1ex]
&=\frac{1}{(2\pi)^3}\int_{\IR^3} \frac{d^3\vec{p}}{\o_{\vec{p}}}\ \frac{\cos(z\vekt{p})}{e^{\bb(q)\vekt{p}}-1}\quad \forall q\in\cO,z\in\IR^4,
\label{eq:regular}
\end{align}
where we write $\o_{\vec{p}}:=\sqrt{\abs{\vec{p}}^{\,2}+m^2}$ and $\vekt{p}=(\o_{\vec{p}},\vec{p})$.

Next, we prove a lemma which yields two useful identities involving derivatives of the inverse-temperature four vector field $\bb$. These identities are imposed by the constraints on the regular part $\mathcal{W}_2$ of the point-split two-point function $\sfw_2$ of $\o$, eqns. \eqref{eq:Wconstr1} and \eqref{eq:Wconstr2}.

\begin{lemma}
\label{lemma:constraints}
Let $\cO\subseteq \cM$ be convex and open, and let $\o$ be an analytic Hadamard state on $\cA_m(\cM)$ which fulfills the $[\bb,\cO]$-LKMS condition. Then for all $q\in\cO$ the following hold:
\begin{align}
\vekt{p}^\mu\vekt{p}^\l \del_\mu^{[q]}\bb_\l(q)&=0,
\label{eq:betaconstr1}\\
(\del_\mu^{[q]} \bb_\kap\vekt{p}^\kap)(\del^\mu_{[q]} \bb_\l \vekt{p}^\l)&=\frac{e^{\bb(q)\vekt{p}}-1}{e^{\bb(q)\vekt{p}}+1}\Box_q \bb_\nu(q)\vekt{p}^\nu,
\label{eq:betaconstr2}
\end{align}
where $\vekt{p}=(\o_{\vec{p}}, \vec{p})$ for $\vec{p}\in\IR^3 $, with $\o_{\vec{p}}=\sqrt{\vec{p}^{\,2}+m^2}$, $m\geq 0$.
\end{lemma}

\begin{Proof}[Proof.] We start with the proof of the first constraint in lemma \ref{lemma:constraints}.Let $\fw_\cO\in\cD'(\cO\times\IR^4)$ be the extension of the point-split variable-two-point function $\sfw_\cO\in\cD'(\cN_0)$ which occurs in Def. \ref{defi:LKMS}. As discussed at the end of the previous subsection, the constraint \eqref{eq:Wconstr1} implies that the regular part $\mathcal{W}_\cO\in C^A(\cO\times\IR^4)$ of $\fw_\cO$, given by eq.\ \eqref{eq:regular}, fulfills 
\ben
\del_{[q]}^\mu \del_\mu^{[z]}\mathcal{W}_\cO(q,z)= 0\quad\forall q\in\cO,z\in\IR^4.
\label{eq:Wextconstr1}
\een
According to standard results on the differentiability of parametric integrals we thus obtain from eq.\ \eqref{eq:regular} that
\ben
\del_\mu^{[q]}\del^\mu_{[z]}\mathcal{W}_\cO(q,z)= \frac{1}{(2\pi)^3}\int_{\IR^3}\frac{d^3\vec{p}}{\o_{\vec{p}}}\frac{e^{\bb(q)\vekt{p}}}{(e^{\bb(q)\vekt{p}}-1)^2}\vekt{p}^\mu\vekt{p}^\l\del_\mu^{[q]}\bb_\l(q)\sin(z\vekt{p})=0\quad\forall q\in\cO,z\in\IR^4,
\label{eq:intconstr1}
\een
We introduce the shorthand notation
\ben
I(\vec{p};q):=\vekt{p}^\mu\vekt{p}^\l \del_\mu^{[q]}\bb_\l(q)\cdot\frac{e^{\bb(q)\vekt{p}}}{\o_{\vec{p}}\left(e^{\bb(q)\vekt{p}}-1\right)^2}.
\label{eq:Idefi}
\een
Clearly, the function $I$ is polynomially bounded and continuous in $\vec{p}\in\IR^3$. It thus gives rise to a a tempered distribution, i.e. $I(\cdot;q)\in\cS'(\IR^3)$ for all $q\in\cO$.

Expanding the sine in the integrand on the left-hand side of eq.\ \eqref{eq:intconstr1} in terms of exponential functions one finds by simple manipulations that eq.\ \eqref{eq:intconstr1} can be rewritten as 
\ben
\frac{1}{2\pi}\int_\IR^3 d^3\vec{p}\left[e^{it\o_{\vec{p}}}I(-\vec{p};q)-e^{-it\o_{\vec{p}}}I(\vec{p};q)\right]e^{-i\vec{p}\vec{z}}=0\quad \forall q\in\cO,z\in\IR^4.
\label{eq:intconstr2}
\een
The left-hand side of the latter equation is (a multiple of) the inverse Fourier transform of the term in square brackets, which is well-defined due to the properties of the tempered distribution $I$. Since the Fourier transform $\cF\from \cS'(\IR^3)\to \cS'(\IR^3)$ is one-to-one it follows from  eq.\ \eqref{eq:intconstr2} that
\ben
I(-\vec{p};q)=e^{2it \o_{\vec{p}}}I(\vec{p};q)\quad \forall \vec{p}\in\IR^3;q\in\cO, t\in\IR.
\label{eq:intconstr3}
\een
In particular, eq.\ \eqref{eq:intconstr3} holds for $t=0$, which implies 
\ben
I(-\vec{p};q)=I(\vec{p};q)\quad \forall \vec{p}\in\IR^3,q\in\cO.
\label{eq:intconstr4}
\een
From the definition of $I$, eq.\ \eqref{eq:Idefi} we immediately find that eq.\ \eqref{eq:intconstr4} is fulfilled if and only if $I$ vanishes identically. Since the second factor in eq.\ \eqref{eq:Idefi} is non-vanishing it follows that
\ben
\vekt{p}^\mu\vekt{p}^\l \del_\mu^{[q]}\bb_\l(q)=0 \quad \forall q\in\cO,\vekt{p}=(\o_{\vec{p}},\vec{p}),\vec{p}\in\IR^3,
\non \een
which proves eq.\ \eqref{eq:betaconstr1}.

In order to prove the second constraint in lemma \ref{lemma:constraints} we notice that, according to eq. \eqref{eq:Wconstr2}, it holds
\ben
\Box_q W_\cO(q,z)=\frac{1}{(2\pi^3)}\int_{\IR^3} \frac{d^3\vec{p}}{\o_{\vec{p}}}\cos(z\vekt{p})\Box_q\left(\frac{1}{e^{\bb(q)\vekt{p}}-1}\right)=0\quad\forall q\in\cO,z\in\IR^4,
\non\een
By similar arguments as above one then shows that this is fulfilled if and only if
\ben
\Box_q\left(\frac{1}{e^{\bb(q)\vekt{p}}-1}\right)=0\quad \forall q\in\cO,\vekt{p}=(\abs{\vec{p}},\vec{p}),\vec{p}\in\IR^3.
\non \een
A straightforward calculation then yields eq.\ \eqref{eq:betaconstr2}. 
\end{Proof}

This lemma is now used in the proof of the following proposition, revealing the spacetime-dependence of the inverse-temperature four-vector field $\bb$ in the region $\cO$. As discussed in the introduction, it is found that the LKMS condition is quite restrictive since the possible inverse-temperature vector field $\bb$ either has a linear dependence on spacetime or is constant. In the massive case only the latter possibility occurs.

\begin{prop}
\label{prop:betaofq}
Let $\cO\subseteq\cM$ be an open neighbourhood, and let $\o$ an analytic Hadamard state on $\cA_m(\cM)$ which fulfills the $[\bb,\cO]$-LKMS condition. Then the following holds:

\vspace{12pt}
\noindent If m=0, there exists a $\tilde{\bb}\in\IR^4$ such that either:
\begin{itemize}
\item[(i)] $\cO\subseteq V_+-\tilde{\vekt{\bb}}$, and 
\ben
\bb(q)=c_\o (q+\tilde{\bb})\quad\forall q\in \cO,
\non \een
where $c_{\o}>0$ is a state dependent constant, or
\item[(ii)] $\cO\subseteq V_--\tilde{\bb}$, and 
\ben
\bb(q)=c_\o (q+\tilde{\bb})\quad\forall q\in \cO,
\non \een
where $c_{\o}<0$ is a state dependent constant, or
\item[(iii)] $\tilde{\bb}\in V_+$ and $\bb(q)=const.=\tilde{\bb}$ for all $q\in\cO$.
\end{itemize}
\noindent If $m>0$, there exists a $\tilde{\bb}\in V_+$ such that $\bb(q)=\text{const.}=\tilde{\bb}$ for all $q\in \cO$.
\end{prop}

We will split the proof of Proposition \ref{prop:betaofq} into two parts: one for the massless case ($m=0$) and one for the massive case ($m>0$). In the proof we will make use of the two following lemmas:

\begin{lemma}
\label{lemma:A0}
Let $\vekt{p}=(\abs{\vec{p}},\vec{p}),\vec{p}\in\IR^3$. A $\binom{0}{2}$-tensor $A_{\mu\nu}$ fulfills 
$\vekt{p}^\mu\vekt{p}^\nu A_{\mu\nu}=0\ \forall \vec{p}\in \IR^3$ if and only if it is of the form
\ben
A_{\mu\nu}=c\cdot\eta_{\mu\nu}+\O_{\mu\nu},
\label{eq:Amunu}
\een
 with $c\in\IR$ and $\O_{\mu\nu}=-\O_{\nu\mu}$.
\end{lemma}

\begin{Proof}[Proof.]
The proof can be found in Appendix A.
\end{Proof}

\begin{lemma}
 \label{lemma:Am}
 Let $m>0$ and $\o_{\vec{p}}:=\sqrt{\vec{p}^{\ 2}+m^2}$ for $\vec{p}\in\IR^3$. A $\binom{0}{2}$-tensor $A_{\mu\nu}$ is antisymmetric, i.e. $A_{\mu\nu}=-A_{\nu\mu}$, if and only if
\ben
\vekt{p}^\mu\vekt{p}^\nu A_{\mu\nu}=0\quad \forall \vekt{p}=(\o_{\vec{p}},\vec{p}),\vec{p}\in \IR^3.
\non \een
 \end{lemma}
 
\begin{Proof}[Proof.]
To the best knowledge of the author this lemma has first been proven in \cite{Hue05}. Since the results of the latter work have not been published, the proof is included in Appendix A for convenience of the reader.
\end{Proof}

We can now give the proof of Proposition \ref{prop:betaofq} for the case of the massless Klein-Gordon field. 

\begin{Proof}[Proof of Proposition \ref{prop:betaofq} for $m=0$.]
 Assume that $\o$ is a $[\bb,\cO]$-LKMS state on $\cA_0(\cM)$. Then the combination of the first constraint in Lemma \ref{lemma:constraints}, eq.\ \eqref{eq:betaconstr1}, and Lemma \ref{lemma:A0} yields
\ben
\del_\mu^{[q]}\bb_\nu(q)=A(q)\eta_{\mu\nu}+\O_{\mu\nu}(q) \quad \forall q\in\cO,
\label{eq:delbetagen}
\een
where $A,\O_{\mu\nu}\in C^1(\cO,\IR)$, with $\O_{\mu\nu}=-\O_{\nu\mu}$. For notational convenience we will from now on supress the argument $q$ in the derivatives, i.e. we simply write $\del_\mu\equiv\del_\mu^{[q]}$. From eq.\ \eqref{eq:delbetagen} we read off
\begin{align*}
A(q)\eta_{\mu\nu}&=\frac{1}{2}\left(\del_\mu\bb_\nu(q)+\del_\nu\bb_\mu(q)\right),\quad \text{and}\\
\O_{\mu\nu}(q)&=\frac{1}{2}\left(\del_\mu\bb_\nu(q)-\del_\nu\bb_\mu(q)\right).
\end{align*}
Using the commutativity of partial derivatives we find that this implies the relation
\begin{align}
 \del_\l \O_{\mu\nu}(q)=(\del_\mu A)(q)\eta_{\l\nu}-(\del_\nu A)(q)\eta_{\l\mu}\ 
\label{eq:delOmega}
\end{align}
We next insert eq.\ \eqref{eq:delbetagen} into the second constraint in Lemma \ref{lemma:constraints}, eq.\ \eqref{eq:betaconstr2}, which gives
\begin{align}
(\del_\mu\bb_\l(q)\vekt{p}^\l)(\del^\mu\bb_\l(q)\vekt{p}^\l)=(\vekt{p}^\kap \O_{\mu\kap}(q))(\vekt{p}^\l\O^\mu_{\ \l}(q))\quad\forall q\in\cO,\vekt{p}=(\abs{\vec{p}},\vec{p}),\vec{p}\in\IR^3,
\label{eq:betaconstr30}
\end{align}
where the remaining terms on the right-hand side vanish since $\vekt{p}=(\abs{\vec{p}},\vec{p})$ is lightlike and $\O_{\mu\nu}(q)$ is antisymmetric. On the other hand, eq.\ \eqref{eq:delbetagen} gives
\begin{align}
\Box \bb_\l(q)\vekt{p}^\l=\vekt{p}^\l\del_\l A(q)+\vekt{p}^\l \del^\mu \O_{\mu_\l}(q)\quad \forall q\in\cO,\vekt{p}=(\abs{\vec{p}},\vec{p}),\vec{p}\in\IR^3.
\label{eq:boxbetagen}
\end{align}
Since it follows from eq.\ \eqref{eq:delOmega} that $\del^\mu \O_{\mu_\l}=-3\del_\l A$ we can further simplify eq.\ \eqref{eq:boxbetagen} to
\ben
\Box \bb_\l(q)\vekt{p}^\l=-2\vekt{p}^\l\del_\l A(q)\quad \forall q\in\cO,\vekt{p}=(\abs{\vec{p}},\vec{p}),\vec{p}\in\IR^3.
\non \een
Plugging this and eq.\ \eqref{eq:betaconstr30} into eq.\ \eqref{eq:betaconstr2} gives
\ben
\vekt{p}^\l \left(\del_\l A(q)+\frac{e^{\bb(q)\vekt{p}}+1}{2(e^{\bb(q)\vekt{p}}-1)}\vekt{p}^\kap \O_{\mu\kap}(q)\O^\mu_{\ \l}(q)\right)=0\quad \forall q\in\cO,\vekt{p}=(\abs{\vec{p}},\vec{p}),\vec{p}\in\IR^3.
\non \een
In particular, we can replace $\vekt{p}$ by $r\vekt{p}=(r\abs{\vec{p}},r\vec{p}),r>0$ in the latter equation and subsequently divide by $r$, which yields
\ben
\vekt{p}^\l\del_\l A(q)=-r\frac{e^{r\bb(q) \vekt{p}}+1}{2(e^{r\bb(q)\vekt{p}}-1)}\vekt{p}^\l\vekt{p}^\kap\O_{\mu\kap}(q)\O^\mu_{\ \l}(q)\quad \forall q\in\cO. 
\non \een
Since the left hand side does not depend on $r$ anymore, we find that the latter equation can be fulfilled if and only if  $\vekt{p}^\l\del_\l A(q)=0$ for all $\vekt{p}$, which implies $\del_\l A(q)=0$ in $\cO$, and thus
\ben
A(q)=\text{const.}=\colon c_\o\quad \forall q\in\cO.
\non \een
This implies immediately by eq.\ \eqref{eq:delOmega} that $\del_\l\O_{\mu\nu}(q)=0$ for all $q\in\cO$, and therefore
\ben
\O_{\mu\nu}(q)=\text{const.}=\colon C_{\mu\nu}\quad \forall q\in\cO.
\non \een
Thus, eq.\ \eqref{eq:delbetagen} reduces to
\ben
\del_\mu \bb_\nu(q)=c_\o\eta_{\mu\nu} +C_{\mu\nu}\quad \forall q\in\cO,
\non \een
with $C_{\mu\nu}=-C_{\nu\mu}$. Integrating the latter we obtain that the inverse-temperature four-vector field $\bb$ is constrained to be of the form
\ben
\bb_\nu(q)=c_\o q_\nu +C_{\mu\nu} q^\mu+\tilde{\bb}_\nu\quad \forall q\in\cO,
\label{eq:betagen}
\een
with $c_\o,C_{\mu\nu}\in\IR,\tilde{\bb}\in\IR^4$, and $C_{\mu\nu}=-C_{\nu\mu}$. Inserting this into eq.\ \eqref{eq:betaconstr2} yields
\begin{align}
0&=(c_\o\vekt{p}_\mu+C_{\mu\nu}\vekt{p}^\nu)(c_\o\vekt{p}^\mu +C^\mu_{\ \nu}\vekt{p}^\nu)\quad \forall \vekt{p}=(\abs{\vec{p}},\vec{p}),\vec{p}\in\IR^3.
\label{eq:Csystem}
\end{align}
Choosing $\vec{p}=(1,0,0)$, $\vec{p}=(0,1,0)$ and $\vec{p}=(0,0,1)$, respectively, writing out the summations and exploiting the antisymmetry of $C_{\mu\nu}$ we find that \eqref{eq:Csystem} reduces to a simple system of linear equations for the components of $C_{\mu\nu}$ which has the the unique solution $C_{\mu\nu}=0$ for all $\mu,\nu\in\{0,1,2,3\}$. Thus, according to eq.\ \eqref{eq:betagen} there are the following alternatives for the inverse temperature four-vector field $\bb$:
\ben
\bb(q)=\begin{cases}c_\o (q+\tilde{\bb})& \ c_\o\neq 0,\\
						\tilde{\bb}&\ c_\o=0.
						\end{cases}
\non \een
with some constant $\tilde{\bb}\in\IR^4$.

Since for any any $q\in\cO$ we have the constraint $\bb(q)\in V_+$, it follows that $q\in\{V_+-\tilde{\bb}\}$ for $c_\o> 0$ resp. $q\in\{V_--\tilde{\bb}\}$ for $c_\o< 0$. If $c_\o=0$ we must have $\tilde{\bb}\in V_+$. 
\end{Proof}

It remains to prove Proposition \ref{prop:betaofq} for the massive Klein-Gordon field.

\begin{Proof}[Proof of Proposition \ref{prop:betaofq} with $m>0$.]
Assume that $\o$ is a $[\bb,\cO]$-LKMS state on $\cA_m(\cM)$ with $m>0$. Combining the first constraint in Lemma \ref{lemma:constraints}, eq.\ \eqref{eq:betaconstr1}, with Lemma \ref{lemma:Am} it follows that  $\del_\mu\bb_\nu(q)$ is antisymmetric, i.e.
\ben
\del_\mu\bb_\nu(q)=-\del_\mu\bb_\nu(q)\quad \forall q\in\cO.
\non \een
By repeated use of this antisymmetry, together with the commutativity of partial derivatives, we find (suppressing again the argument $q$ in the derivatives)
\begin{align*}
\del_\l \del_\mu \bb_\nu(q)=-\del_\l \del_\mu \bb_\nu(q)\quad \forall q\in\cO,
\end{align*}
which implies $\del_\l \del_\mu \bb_\nu(q)=0$ for all $q\in\cO$. Thus, for $\mu,\nu\in\{0,1,2,3\}$ there are constants $C_{\mu\nu}\in\IR$ such that
\ben
\del_\mu\bb_\nu(q)= C_{\mu\nu}\quad \forall q\in\cO.
\label{eq:delbetaconst}
\een
Next, by Lemma \ref{lemma:constraints} we have the following identity:
\ben
(e^{\bb(q)\vekt{p}}+1)(\del_\mu \bb_\kap\vekt{p}^\kap)(\del^\mu \bb_\l \vekt{p}^\l)=(e^{\bb(q)\vekt{p}}-1)\Box_q \bb_\nu(q)\vekt{p}^\nu\quad \forall q\in\cO,\vekt{p}=(\o_{\vec{p}},\vec{p}),\vec{p}\in\IR^3.
\non \een
Inserting eq.\ \eqref{eq:delbetaconst}, which in particular implies $\Box_q \bb_\l(q)=0$, into eq.\ \eqref{eq:betaconstr2} gives
\ben
\left(C_{\mu\kap}\vekt{p}^\kap\right)\left(C^\mu_{\ \l}\vekt{p}^\l\right)=0\quad \forall \vekt{p}=(\o_{\vec{p}},\vec{p}),\vec{p}\in\IR^3.
\label{eq:Csystemm}
\een
For $\vec{p}=(0,0,0)$ eq.\ \eqref{eq:Csystemm} reduces to
\ben
0=m^2(C_{10}^2+C_{20}^2+C_{30}^2)
\non \een
and it follows that $C_{\mu 0}=0$ for all $\mu\in\{0,1,2,3\}$ (note that $C_{00}=0$ due to the antisymmetry of $C_{\mu\nu}$). Now we choose $\vec{p}=(1,0,0)$ in eq.\ \eqref{eq:Csystemm}, which gives
\ben
0=C_{21}^2+C_{31}^2\quad\Leftrightarrow\quad C_{21}=C_{31}=0.
\non \een
Analogously, choosing $\vec{p}=(0,1,0)$ in eq.\ \eqref{eq:Csystemm} we find $C_{32}=0$. This, together with the results obtained before and the antisymmetry of $C_{\mu\nu}$ finally yields $C_{\mu\nu}=0$ for all $\mu,\nu\in\{0,1,2,3\}$, i.e. we have by eq.\ \eqref{eq:delbetaconst}
\ben
\del_\mu\bb_\nu(q)=0\quad\forall q\in\cO.
\non \een
Integrating the latter equation we obtain that the inverse-temperature four-vector field $\bb$ has to be of the form
\ben
\bb_\nu(q)=\tilde{\bb}_\nu\quad \forall q\in\cO,
\non \een
for some constant $\tilde{\bb}\in V_+$. This completes the proof of Proposition \ref{prop:betaofq} in the case $m>0$.
 \end{Proof}

Now, the proof of Theorem \ref{thm:mainresult} can be completed:

\begin{Proof}[Proof of Theorem \ref{thm:mainresult}.]
Since the class of $[\bb,\cO]$-LKMS states fulfills the analytic Hadamard condition by assumption, we can proceed by the following chain of arguments: We first note that the $[\bb,\cO]$-LKMS condition implies the $[\bb,\cM]$-LKMS condition for any open neighbourhood $\cO\subseteq\cM$, with the spacetime-dependence of $\bb$ found in Prop. \ref{prop:betaofq}. This follows via analytic continuation (in the position variable $q$) of the regular part $\mathcal{W}_\cO\in C^A(\cN_\cO)$  of the point-split two-point function ${\mathsf w}_\cO\in\cD'(\cN_\cO)$ of $\o$ in $\cO$. That this can be done is easily read off from the analytic wave front set $\WF_A({\mathsf w}_\cO)$, given by eq.\ \eqref{eq:WFAw} in Lemma \ref{lemma:pullback}. The same argument is then used to extend the point-split two-point function either to the whole lightcone in the case of varying $\bb$, or to all of Minkowski spacetime $\IM$ in the case of constant $\bb$. This completes the proof.
\end{Proof}

\section{Outlook}

Although we restricted ourselves to the simple case of a free scalar field on Minkowski spacetime the ingredients and methods used in the analysis (analytic wavefront sets, pullback of distributions under diffeomorphisms) are of a fairly general nature. In particular our investigation demonstrates that these techniques and methods may also be of use in the realm of thermal physics and quantum statistical mechanics. It seems to be an interesting feature of the LKMS condition that once it has been given a precise formulation in terms of these ingredients, including the analytic Hadamard condition, one immediately obtains a closed expression for the two-point function $\o_2$ and the further calculations are rather straightforward applications of linear algebra.

In view of results known for LTE states of the free massless Dirac field \cite{Bah06} it seems plausible that results similar to those obtained in \cite{GPV15} and in the present work also hold in the case of free quantum fields of higher spin. Moreover, since the notions of microlocal analysis of distributions generalize immediately to arbitrary manifolds a generalization of the LKMS condition to curved spacetimes (at least to such that are equipped with an analytic metric) seems to be within reach. It is an highly interesting question if quantum states of ``hot-bang type'' also exist for the massless Klein-Gordon field in stationary spacetimes or if they are just a mathematical artifact arising from the unusually high symmetry of Minkowski spacetime together with the conformal symmetry of the massless Klein-Gordon field. We hope to come back to this circle of questions elsewhere.

\subsubsection*{Acknowledgements}
The author likes to thank Rainer Verch, Ko Sanders and Steffen Pottel for helpful discussions and Tim Laux for valuable comments on an earlier version of the manuscript. Financial support by the International Max Planck Research School ``Mathematics in the Sciences'' of the MPI MIS is gratefully acknowledged.

\appendix
\section{Proof of two lemmas}

In this appendix we give the proofs of Lemma \ref{lemma:A0} and Lemma \ref{lemma:Am}.

\begin{Proof}[Proof of Lemma \ref{lemma:A0}.]
If  $A_{\mu\nu}$ is of the form \eqref{eq:Amunu}, it immediately follows that $\vekt{p}^\mu\vekt{p}^\nu A_{\mu\nu}=0\ \forall \vec{p}\in \IR^3$ by the antisymmetry of $\O_{\mu\nu}$ and the fact that $\vekt{p}$ is lightlike. 

Assume conversely that $\vekt{p}^\mu\vekt{p}^\nu A_{\mu\nu}=0\ \forall \vec{p}\in \IR^3$. Decomposing $A_{\mu\nu}$ into its symmetric part $S_{\mu\nu}$ and its antisymmetric part $\O_{\mu\nu}$ yields
\ben
0=\vekt{p}^\mu\vekt{p}^\nu S_{\mu\nu}+\underbrace{\vekt{p}^\mu\vekt{p}^\nu \O_{\mu\nu}}_{=0}=\vekt{p}^\mu\vekt{p}^\nu S_{\mu\nu}.
\non \een
Using Einstein's summation convention for spatial indices $i,j,k\in\{1,2,3\}$, this can be written as
\ben
0=\abs{\vec{p}}^2 S_{00}+2\abs{\vec{p}}\vec{p}^{\,i} S_{0 i} +\vec{p}^{\,j}\vec{p}^{\,k} S_{jk}\quad \forall \vec{p}\in\IR^3,
\label{eq:symvan0}
\een
where we have used the symmetry of $S_{\mu\nu}$. In particular we can replace $\vec{p}$ by $-\vec{p}$ in the latter equation, which implies
\ben
0=\abs{\vec{p}}\vec{p}^{\,i} S_{0 i}\quad \forall \vec{p}\in\IR^3.
\non \een
This can be fulfilled  for all $\vec{p}\in\IR^3$ if and only if $S_{0i}=0=S_{i0}$ $\forall i\in\{1,2,3\}$. Therefore, by eq.\ \eqref{eq:symvan0} we have
\ben
0=\abs{\vec{p}}^2 S_{00}+\vec{p}^{\,j} \vec{p}^{\,k} S_{jk}\quad \forall\vec{p}\in\IR^3.
\non \een
Since the right-hand side of the latter equation is a polynomial (of second order) in the components of $\vec{p}$, we find that the coefficients of this polynomial have to vanish, which implies $S_{00}=-S_{ii}=:c$ and $S_{jk}=0$ for all $j\neq k\in\{1,2,3\}$.  This together with $S_{0i}=S_{i0}=0$ finally yields $S_{\mu\nu}=c\cdot\eta_{\mu\nu}$, which completes the proof.
\end{Proof}

\begin{Proof}[Proof of Lemma \ref{lemma:Am}.]
Assume first that $A_{\mu\nu}=-A_{\nu\mu}$. Then clearly $\vekt{p}^\mu\vekt{p}^\nu A_{\mu\nu}=0$ for all $\vekt{p}=(\o_{\vec{p}},\vec{p}),\vec{p}\in\IR^3$.

Assume conversely that $\vekt{p}^\mu\vekt{p}^\nu A_{\mu\nu}=0$ for all $\vekt{p}=(\o_{\vec{p}},\vec{p}),\vec{p}\in\IR^3$. Decomposing $A_{\mu\nu}$ into its symmetric part $S_{\mu\nu}$ and its antisymmetric part $\O_{\mu\nu}$ then yields 
\ben
0=\vekt{p}^\mu\vekt{p}^\nu S_{\mu\nu}+\underbrace{\vekt{p}^\mu\vekt{p}^\nu\O_{\mu\nu}}_{=0}\quad \forall\vekt{p}=(\o_{\vec{p}},\vec{p}),\vec{p}\in\IR^3.
\non \een
where the second term on the right-hand side vanishes due to the antisymmetry of $\O_{\mu\nu}$.  By the symmetry of $S_{\mu\nu}$ we thus find (using Einstein summation for spatial indices $i,j,k\in\{1,2,3\}$):
\ben
0=\o_{\vec{p}}^2\, S_{00}+2 \o_{\vec{p}}\,{\vec{p}}^{\,i} S_{0i}+{\vec{p}}^{\,j}{\vec{p}}^{\,k} S_{jk}\quad \forall \vec{p}\in\IR^3.
\label{eq:symvan1}
\een
We may replace $\vec{p}$ by $-\vec{p}$ in the latter equation which implies
\ben
0=\o_{\vec{p}}\,{\vec{p}}^{\,i} S_{0i}
\non \een
which can be fulfilled if and only if $S_{0i}=S_{i0}=0$ for all $i\in\{1,2,3\}$. Hence, it follows from eq.\ \eqref{eq:symvan1} that
\ben
0=\o_{\vec{p}}^2\, S_{00}+{\vec{p}}^{\,j}{\vec{p}}^{\,k} S_{jk}\quad \forall \vec{p}\in\IR^3.
\label{eq:symvan2}
\een
In particular we can choose $\vec{p}=0$, i.e. $\o_{\vec{p}}=m^2$, which gives $0=m^2 S_{00}$. Since by assumption $m>0$ it follows that $S_{00}=0$, and eq.\ \eqref{eq:symvan2} reduces to
\ben
0={\vec{p}}^{\,j}{\vec{p}}^{\,k} S_{jk}\quad \forall \vec{p}\in\IR^3.
\non \een
Again, this can be fulfilled if and only if $S_{jk}=0$ for $j,k\in\{1,2,3\}$. This, together with $S_{00}=0$ and $S_{0i}=S_{i0}=0$ finally yields $S_{\mu\nu}=0$, which completes the proof.
\end{Proof}

\bibliographystyle{amsplain}

\begin{thebibliography}{10}

\bibitem{Bah06}
B.~Bahr, \emph{{The hot bang state of massless fermions}}, Lett. Math. Phys.
  \textbf{78} (2006), 39--54.

\bibitem{BR97}
O.~Bratteli and D.W. Robinson, \emph{{Operator Algebras and Quantum Statistical
  Mechanics 2, Second Edition}}, Springer, 1997.

\bibitem{BB94}
J.~Bros and D.~Buchholz, \emph{{Towards a relativistic KMS-condition}},
  Nucl.Phys. \textbf{B429} (1994), 291--318.

\bibitem{BB96}
J.~Bros and D.~Buchholz, \emph{{Axiomatic analyticity properties and representations of
  particles in thermal quantum field theory}}, Annales Poincare Phys.Theor.
  \textbf{64} (1996), 495--522.

\bibitem{Buc03}
D.~Buchholz, \emph{{On hot bangs and the arrow of time in relativistic quantum
  field theory}}, Commun.Math.Phys. \textbf{237} (2003), 271--288.

\bibitem{BOR02}
D.~Buchholz, I.~Ojima, and H.~Roos, \emph{{Thermodynamic properties of
  non-equilibrium states in quantum field theory}}, Annals Phys. \textbf{297}
  (2002), 219--242.

\bibitem{DHP11}
T.-P.~Hack C.~Dappiaggi and N.~Pinamonti, \emph{{Approximate KMS states for
  scalar and spinor fields in Friedmann-Robertson-Walker spacetimes}}, Annales
  Henri Poincare \textbf{12} (2011), 1449--1489.

\bibitem{CVJ03}
J.~Casas-V\'asquez and D.~Jou, \emph{{Temperature in non-equilibrium states: A
  review of open problems and current proposals}}, Rep. Prog. Phys. \textbf{66}
  (2003), 1937.

\bibitem{Dix78}
W.G. Dixon, \emph{{Special Relativity}}, Cambridge University Press, 1978.

\bibitem{FV13}
C.J. Fewster and R.~Verch, \emph{{The Necessity of the Hadamard Condition}},
  Class. Quant. Grav. \textbf{30} (2013), 235027.

\bibitem{Gra16}
M.~Gransee, \emph{{Local Thermal Equilibrium States in Relativistic Quantum
  Field Theory}}, in: Quantum Mathematical Physics: A Bridge between
  Mathematics and Physics (eds.: F. Finster et al.) (2016), 101--117.

\bibitem{GPV15}
M.~Gransee, N.~Pinamonti, and R.~Verch, \emph{{KMS-like properties of local
  thermal equilibrium states in quantum field theory}}, arXiv:1508.05585
  [math-ph] (2015).

\bibitem{Haa92}
R.~Haag, \emph{{Local Quantum Physics}}, Springer, 1992.

\bibitem{HHW67}
R.~Haag, N.M. Hugenholtz, and M.~Winnink, \emph{{On the equilibrium states in
  quantum statistical mechanics}}, Commun.Math.Phys. \textbf{5} (1967),
  215--236.

\bibitem{HTP77}
R.~Haag and E.~Trych-Pohlmeyer, \emph{{Stability properties of equilibrium
  states}}, Commun. Math. Phys. \textbf{56} (1977), 213.

\bibitem{Hac15}
T.-P. Hack, \emph{{Cosmological Applications of Algebraic Quantum Field Theory
  in Curved Spacetimes}}, {SpringerBriefs in Mathematical Physics} \textbf{6}
  (2015).

\bibitem{HL16}
S.~Hollands and R.~Longo, \emph{{Non-equilibrium thermodynamics in conformal
  field theory}}, {arXiv:1605.01581 [hep-th]} (2016).

\bibitem{Hoe90}
L.~H{\"o}rmander, \emph{{The Analysis of Linear Partial Differential Operators
  I, 2nd edition}}, Springer, 1990.

\bibitem{Hue05}
R.~H{\"u}bener, \emph{{ Lokale Gleichgewichtszust{\"a}nde massiver Bosonen}},
  Diploma Thesis, Universit{\"a}t G{\"o}ttingen (2005).

\bibitem{PW78}
W.~Pusz and S.L. Woronowicz, \emph{{Passive states and KMS states for general
  quantum systems}}, Commun. Math. Phys. \textbf{58} (1978), 273--290.

\bibitem{Rad96}
M.~J. Radzikowski, \emph{{Micro-local approach to the Hadamard condition in
  quantum field theory on curved space-time}}, Commun. Math. Phys. \textbf{179}
  (1996), 529--553.

\bibitem{RS80}
M.~Reed and B.~Simon, \emph{{Methods of Modern Mathematical Physics I:
  Functional Analysis (revised and enlarged edition)}}, Elsevier, 1980.

\bibitem{Rue00}
D.~Ruelle, \emph{{Natural nonequilibrium states in quantum statistical
  mechanics}}, J. Stat. Phys. \textbf{98} (2000), 57--75.

\bibitem{SV08}
J.~Schlemmer and R.~Verch, \emph{{Local thermal equilibrium states and quantum
  energy inequalities}}, Annales Henri Poincare \textbf{9} (2008), 945--978.

\bibitem{Sew13}
G.L. Sewell, \emph{{Local thermodynamic equilibrium at three levels}}, {Rep.
  Math. Phys.} \textbf{72} (2013), no.~3, 389--404.

\bibitem{SVW02}
A.~Strohmaier, R.~Verch, and M.~Wollenberg, \emph{{Microlocal analysis of
  quantum fields on curved space-times: Analytic wavefront sets and
  Reeh-Schlieder theorems}}, J.Math.Phys. \textbf{43} (2002), 5514--5530.

\bibitem{Ver12}
R.~Verch, \emph{{Local covariance, renormalization ambiguity, and local thermal
  equilibrium in cosmology}}, in "Quantum Field Theory and Gravity", F. Finster
  et al. (eds.), Birkh\"auser (2012), 229--256.

\bibitem{Wal94}
R.~M. Wald, \emph{{Quantum Field Theory in Curved Spacetime and Black Hole
  Thermodynamics}}, The University Of Chicago Press, 1994.

\end{thebibliography}
\providecommand{\bysame}{\leavevmode\hbox to3em{\hrulefill}\thinspace}
\providecommand{\MR}{\relax\ifhmode\unskip\space\fi MR }
\providecommand{\href}[2]{#2}

\end{document}